\documentclass[aps,prl,twocolumn,superscriptaddress,showpacs]{revtex4}
\usepackage{graphicx}

\begin{document}


\title{Absence of magnetic order in Yb$_3$Ga$_5$O$_{12}$: relation between
phase transition and entropy in geometrically frustrated materials}


\author{P. \surname{Dalmas de R\'eotier}}
\affiliation{Commissariat \`a l'Energie Atomique, D\'epartement de Recherche
Fondamentale sur la Mati\`ere Condens\'ee\\
F-38054 Grenoble Cedex 9, France}
\author{A. Yaouanc}
\affiliation{Commissariat \`a l'Energie Atomique, D\'epartement de Recherche
Fondamentale sur la Mati\`ere Condens\'ee\\
F-38054 Grenoble Cedex 9, France}
\author{P.C.M. Gubbens}
\affiliation{Interfacultair Reactor Instituut, 2629 JB Delft, The Netherlands}
\author{C.T. Kaiser}
\affiliation{Interfacultair Reactor Instituut, 2629 JB Delft, The Netherlands}
\author{C. Baines}
\affiliation{Laboratory for Muon Spectroscopy, Paul Scherrer Institute, 
5232 Villigen-PSI, Switzerland}
\author{P.J.C. King}
\affiliation{ISIS Facility, Rutherford Appleton Laboratory, Chilton, Didcot, OX11 0QX, UK}

\date{\today}

\begin{abstract}
From muon spin relaxation spectroscopy experiments, we show 
that the sharp peak ($\lambda$ type anomaly) detected 
by specific heat measurements at 54 mK for the ytterbium gallium garnet 
compound, Yb$_3$Ga$_5$O$_{12}$, does not correspond to the onset of a magnetic
phase transition, but to a pronounced building up of 
dynamical magnetic pair-correlations.
Beside the $\lambda$ anomaly, a broad hump is observed at higher 
temperature in the specific heat of this garnet and other geometrically 
frustrated compounds.
Comparing with other frustrated magnetic systems we infer that a ground
state with long-range order is reached only when at least $1/4-1/3$  
of the magnetic entropy is released at the $\lambda$ transition.
\end{abstract}


\maketitle

A magnetic phase transition to a long-range order occurs at low temperature
for most crystallographically ordered compounds containing a three dimensional
lattice of magnetic ions. Its classical signature is a sharp anomaly in 
the magnetic specific heat (SH) at temperature $T_\lambda$ which corresponds to 
the phase transition temperature; see e.g. 
Ref.~\cite{Morrish65}. In this letter, for convenience, we label such anomaly 
a $\lambda$-anomaly although it may not have all
the properties attributed to such an anomaly \cite{Stanley71}.

However, for some particular lattice structures, the long-range magnetic 
ordering of the magnetic ions may be impeded by their geometric arrangement
which gives rise to frustration of their magnetic interactions 
\cite{Villain79,Diep94}. Most of the experimental studies focus nowadays on 
pyrochlore and gallium garnet compounds, R$_2$T$_2$O$_7$ and 
R$_3$Ga$_5$O$_{12}$ respectively. R denotes a rare earth atom 
and T a transition element. The R ions are arranged on a motif of corner 
sharing tetrahedra 
for the pyrochlore structure. In the garnet case, the R atoms form two 
interpenetrating, non-coplanar, corner sharing triangular 
sublattices. For an experimental review, see e.g. Ref.~\cite{Ramirez01}. 

The results from SH measurements are particularly intriguing. Although 
$|\theta_{\rm CW}|/T_\lambda \gtrsim 1$, where $\theta_{\rm CW}$ is the 
Curie-Weiss 
temperature, long-range order may not be present below $T_\lambda$.

For example, let us review data on pyrochlore compounds. According to SH, 
Gd$_2$Ti$_2$O$_7$ has two magnetic phase transitions at 1 K and $\sim$ 0.75 K 
and no defined structure in the SH data is observed above 1 K \cite{Ramirez02}. 
The detection of magnetic Bragg reflections from neutron scattering at 
50 mK shows that a magnetic structure is 
established \cite{Champion01}. Similarly, Gd$_2$Sn$_2$O$_7$ has a large 
$\lambda$-anomaly in SH and M\"ossbauer spectroscopy \cite{Bonville03}
and muon spin relaxation ($\mu$SR) \cite{Yaouanc04} are consistent with 
the presence of long-range ordering. A $\lambda$-anomaly is detected at 
$\sim$ 1.2 K for Er$_2$Ti$_2$O$_7$ \cite{Blote69} and  long-range order is 
observed at low temperature \cite{Champion03}.
For Yb$_2$Ti$_2$O$_7$ a $\lambda$-anomaly is found at 
$T_\lambda \simeq 0.21$ K \cite{Blote69} but there are no long-range 
magnetic correlations 
below $T_\lambda$ but rather dynamical hysteretic short-range 
correlations \cite{Hodges02}. 
We note that beside the $\lambda$ peak, a broad peak centered near 2~K is 
present in SH.
In the popular spin-ice systems Ho$_2$Ti$_2$O$_7$ and 
Dy$_2$Ti$_2$O$_7$, only a broad anomaly is present in SH \cite{Ramirez99}
and at low temperature the magnetic moment are frozen with no long-range
order \cite{Harris97,Bramwell01}. 

Concerning the garnets, only a broad SH peak is detected at low temperature 
for Gd$_3$Ga$_5$O$_{12}$ \cite{Schiffer94}. This compound does not display any
long-range order \cite{Petrenko98}. 
Dy$_3$Ga$_5$O$_{12}$ displays an SH anomaly 
at $T_\lambda \simeq 0.37$ K, overlapping with a broad peak \cite{Filippi80b}. 
Neutron scattering results show the presence of a
long-range magnetic order below $T_\lambda$ \cite{Filippi81}. Finally, 
while a $\lambda$-anomaly is found for Yb$_3$Ga$_5$O$_{12}$ at 
$T_\lambda$ = 54 mK \cite{Filippi80a}, no information is 
available in the literature about the existence of a short or long-range 
magnetic order.
We note that a relatively large SH hump is also present, centered at $\sim $ 
0.2 K. 

Since no $\lambda$-anomaly is present in the spin-ice systems and in
Gd$_3$Ga$_5$O$_{12}$ we shall not consider these systems any longer.
To gain further understanding of the relationship between entropy and magnetic 
correlations in Yb$_3$Ga$_5$O$_{12}$, we performed $\mu$SR
measurements on this system. Here we assume, that
the measured SH corresponds to magnetic dipole moment degrees of freedom and 
not to another exotic order parameter \cite{Paixao02}.

In the garnet lattice (space group $Ia\bar{3}d$), the crystal field acts on 
the $^2{\rm F}_{7/2}$ state of a Yb$^{3+}$ ion to leave a well isolated 
ground state Kramers doublet and three closely grouped excited doublets 
having an average 
energy corresponding to $\Delta_{\rm ave}/k_{\rm B} \simeq$ 850 K 
\cite{Buchanan67}. In a 
good approximation, the ground state of Yb$^{3+}$ ions diluted in 
Y$_3$Ga$_5$O$_{12}$ is described by an effective spin ${\rm S}^\prime$ = 1/2 
with an isotropic $g$-factor, $g=3.43$ \cite{Carson60,Wolf62}. The same 
description of Yb$^{3+}$ in Yb$_3$Ga$_5$O$_{12}$ is expected to be valid. 
This is supported by the electronic entropy variation below 1 K which is very 
close to $R \ln 2$ \cite{Filippi80a}, as expected for a doublet. 
$R$ is the gas constant. From the measured $g$-factor and bulk magnetization 
\cite{Filippi80a}, the mean value of the Yb$^{3+}$ magnetic moment at low 
temperature is found to be $\sim 1.7 \, \mu_{\rm B}$. 
The interionic interaction is essentially antiferromagnetic
as shown from the susceptibility above $T_\lambda$ which is lower than that
extrapolated from high temperature for a Curie-Weiss behaviour 
\cite{Filippi80a}. From the position of the maximum of the bump of the SH 
an estimate for this interaction strength corresponds to at least 0.2 K. 
The antiferromagnetic nature of the interactions together with
the crystallographic structure lead to frustration.

A sample was prepared by heating the constituent oxides to 
1100$^\circ$ C four times with intermediate grindings. Its quality was 
checked by x-ray diffraction and susceptibility measurements.
Zero-field $\mu$SR measurements were performed at the ISIS and 
PSI muon facilities covering the temperature range from 21 mK to 
290 K. Additional spectra were recorded with a longitudinal field.

\begin{figure}
\includegraphics[scale=0.8]{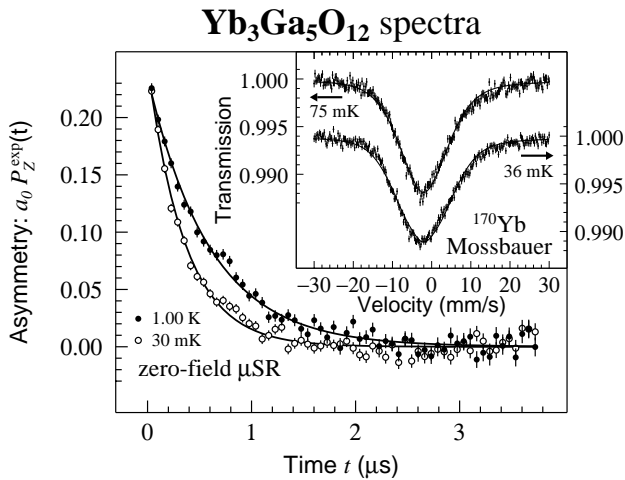}
\caption{Zero-field $\mu$SR spectra 
recorded for Yb$_3$Ga$_5$O$_{12}$, both above and below $T_\lambda$. The solid
lines in the $\mu$SR spectra are the results of fits to exponential relaxation 
functions as explained in the text. 
In the inset $^{170}$Yb M\"ossbauer reproduced from Ref. \cite{Hodges03}. 
These data show the absence of 
long or short-range correlations below $T_\lambda$.}
\label{spectra}
\end{figure}

$^{170}$Yb M\"ossbauer spectroscopy experiments \cite{Hodges03} performed 
between 36 mK and 4.2 K show no resolved hyperfine structure at any 
temperature including below $T_\lambda$ as depicted in the inset of 
Fig.~\ref{spectra}. 
This means that the moments fluctuate at a frequency larger than $\sim$ 300 
$\mu$s$^{-1}$ down to 36 mK. This also indicates the absence of magnetic 
correlations either short- or long-range.

We now present zero-field $\mu$SR results.
The technique consists of implanting polarized (along direction $Z$) muons 
into a 
specimen and monitoring  $P_Z^{\rm exp}(t)$ which is the evolution of the muon 
ensemble polarization 
projected onto direction $Z$ \cite{Dalmas97}. The quantity 
actually measured is the so-called asymmetry corresponding to $a_0 P_Z^{\rm exp}(t)$, 
where $a_0$ $\simeq$ 0.24

For a paramagnet, $P_Z^{\rm exp}(t)$ tracks the dynamics of the magnetic field 
at the muon site, ${\bf B}_{\rm loc}$, reflecting the dynamics of the electronic moments.
In the fast fluctuation or motional narrowing limit, $P_Z^{\rm exp}(t)$ 
takes an exponential form characterized by a relaxation rate, $\lambda_Z$.
 A stochastic approach to 
relaxation in zero-field leads to the relation 
$\lambda_Z = 2 \gamma_\mu^2\Delta^2_{\rm ZF} \tau_c$, assuming a single 
correlation time, $\tau_c$, for the magnetic moment dynamics. $\Delta_{\rm ZF}$ is the 
standard deviation of the components of ${\bf B}_{\rm loc}$ and
$\gamma_\mu$ the muon gyromagnetic 
ratio ($\gamma_\mu$ = 851.615 Mrad s$^{-1}$ T$^{-1}$). 
The motional narrowing limit is valid if $\gamma_\mu\Delta_{\rm ZF} \, \tau_c \ll 1$.

\begin{figure}
\includegraphics[scale=0.8]{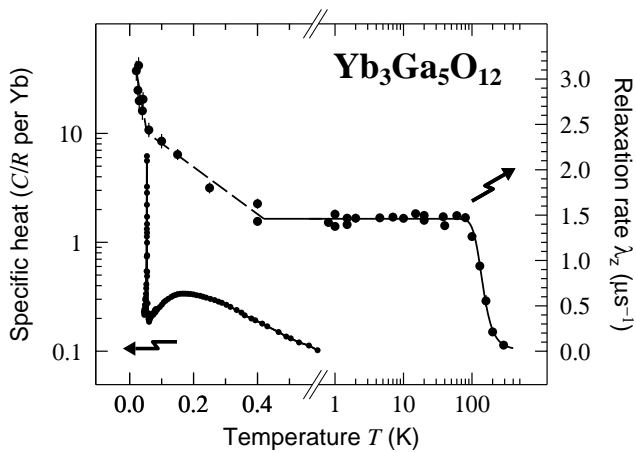}
\caption{Zero-field muon spin-lattice relaxation rate, $\lambda_Z$, versus 
temperature measured for Yb$_3$Ga$_5$O$_{12}$. The solid line is the result of 
a fit to a model explained in the main text. The two straight dashed lines 
for $T \leq 0.4$ K down to 21 mK are guides to the eyes. The specific heat 
(from Ref.~\protect {\cite{Filippi80a}}) of Yb$_3$Ga$_5$O$_{12}$ is also 
reproduced. A marked change of slope in $\lambda_Z(T)$ occurs at 
$T_\lambda$ = 54 mK.}
\label{sh_lambda}
\end{figure}

Two examples of $\mu$SR spectra are shown in Fig.~\ref{spectra}. 
All the spectra were satisfactorily fitted to an exponential function.
The temperature dependence of $\lambda_Z$ is shown in Fig. \ref{sh_lambda}.
Since there is no qualitative change in $P_Z^{\rm exp}(t)$ and since 
$\lambda_Z$ increases monotonously as the temperature is lowered,
there is neither long-range nor short-range magnetic order in 
Yb$_3$Ga$_5$O$_{12}$. This means that the specific heat anomaly at 
$T_\lambda$ does not correspond to the onset of a conventional phase 
transition, in agreement with the $^{170}$Yb M\"ossbauer results. This is
our first result.

The fact that we observe an exponential relaxation function 
below $T_\lambda$ is another important point. It implies that we are in
the fast fluctuation limit (we shall check this quantitatively below) 
and therefore the Yb$^{3+}$ moments continue to fluctuate rapidly 
down to the lowest temperature investigated. This behaviour is very 
different from that in Yb$_2$Ti$_2$O$_7$ where the moments abruptly slow
down below $T_\lambda$ where they are quasi-static \cite{Hodges02}.

We now comment and interpret the shape of $\lambda_Z(T)$.

As for nuclear magnetic resonance \cite{Moriya62}, $\lambda_Z$ can be 
expressed in terms of the static wavevector dependent 
susceptibility, $\chi ({\bf Q})$, and
linewidth of the quasielastic peak of the imaginary component of the dynamical 
susceptibility, $\Gamma({\bf Q})$. These two functions are assumed to be scalar,
consistent with the quasi-isotropic nature of the Kramers ground-state doublet. 
Following Ref. \cite{Yaouanc93} we write
\begin{eqnarray}
\lambda_Z = {\mu_0 \gamma^2_\mu \over 16 \pi^2} \,k_{\rm B} T
\int^{ }_{ } {\cal C}({\bf Q}){\chi({\bf Q}) \over \Gamma({\bf Q})} 
{ {\rm d}^3 {\bf Q} \over( 2\pi)^ 3}.
\label{lambdaZ}
\end{eqnarray}
$\mu_0$ is the permeability of free space and $k_{\rm B}$ the Boltzmann constant.
The term ${\cal C}({\bf Q})$ 
accounts for the interaction of the muon spin with the Yb$^{3+}$ effective spins. 
The integration is over the first Brillouin zone. 

For a Heisenberg magnet at high temperature, {\sl i.e.} if the thermal 
energy is larger than the exchange energy, the ${\bf Q}$ dependence of 
$\chi({\bf Q})$ becomes negligible (see e.g. Ref.~\cite{Lovesey84}) 
and it is simply given by the Curie law. Since $\Gamma({\bf Q})$ is 
temperature independent in the same limit \cite{Lovesey84}, we 
deduce that $\lambda_Z$ should be temperature independent. This is effectively
observed for $0.4 \leq T \leq 80$ K. 

However, above 80 K, $\lambda_Z(T)$ decreases steadily as the sample is heated.
Such a behaviour is sometimes associated with muon diffusion, but it would
be quite unusual in an insulating oxide. In fact the decrease of $\lambda_Z$ 
is due to the relaxation of the Yb$^{3+}$ magnetic moments resulting 
from an Orbach process, i.e. a two-phonon real process with an excited 
crystal-field level as intermediate \cite{Orbach61,Orbachnote}. A global fit 
is achieved with the formula
$\lambda_Z^{-1} = A + B_{\rm me} \exp\left [-{\Delta_e / \left (k_{\rm B} 
T \right )} \right ]$.
$\Delta_e$ is the energy of the excited crystal-field level involved. 
$A \simeq \lambda_Z^{-1}$ when $\lambda_Z$ saturates at low temperature 
(but still $T > 0.4$ K). The constant $B_{\rm me}$ 
models the magneto-elastic coupling of 
the Yb$^{3+}$ spin with the phonon bath. The result of the fit  
for $0.4 \leq T \leq 290$ K is shown in Fig.~\ref{sh_lambda}. Taking 
$\Delta_e/k_{\rm B}$ = $\Delta_{\rm ave}/ k_{\rm B}$ = 850 K, we get 
$A$ = 0.69 (2) $\mu$s and $B_{\rm me}$ = 250 (70) $\mu$s. 
Combining the value of $\lambda_Z$ in the range 0.4 $< T < $ 80 K 
and the value $\tau_c \sim$  38 ps obtained from perturbed angular correlation 
and 
M\"ossbauer data in the same temperature interval \cite{Hodges03}, we compute 
$\Delta_{\rm ZF} \simeq $ 0.16~T which is of the expected magnitude. For 
comparison $\Delta_{\rm ZF} \simeq $ 0.08 T was found for Yb$_2$Ti$_2$O$_7$
above $T_\lambda$ \cite{Yaouanc03}.
  
Information relevant to the influence of the frustrated nature of the magnetic
interactions in Yb$_3$Ga$_5$O$_{12}$ is obtained from 
$\lambda_Z(T)$ at low temperatures. As shown in Fig.~\ref{sh_lambda},
$\lambda_Z$ starts to deviate at $\sim$ 0.4 K from the behaviour expected for 
the Heisenberg Hamiltonian in the high-temperature limit. As we cool down the 
sample, we first observe a mild linear increase of $\lambda_Z$, 
with slope $-2.6\ (3)$ $\mu$s$^{-1}$K$^{-1}$, followed by a sharp increase 
below $T_\lambda$ with slope $-21\ (7) $ $\mu$s$^{-1}$K$^{-1}$.

In the temperature range $T_\lambda < T < 0.4$ K, we note that the broad SH 
peak centered at $\sim$ 0.2 K covers the temperature r\'egime where 
$\lambda_Z(T)$ varies slowly, suggesting that both features reflect the
same physics. Since there are no excited crystal-field energy levels at low 
energy, the hump can not be of a Schottky-type. 
This extra SH reflects short-range 
correlations among groups of spins, see e.g. Ref.~\cite{Morrish65}. This means 
that the spin dynamics is wavevector dependent below $\sim$ 0.4 K. This 
dependence is expected to generate a mild increase of $\lambda_Z$ as shown by 
the explicit computation of Paja and coworkers \cite{Paja02} for a simple 
model.

Now we turn our attention to the region $T < T_\lambda$.
Being in an interstitial site, the muon spin can be  
strongly influenced by magnetic pair-correlations. In contrast, since a 
$^{170}$Yb 
M\"ossbauer nucleus is embedded in a Yb atom which is magnetic, it is mainly 
sensitive to the self correlations. Taking into account that the dynamics 
measured by M\"ossbauer spectroscopy is not wildly different above and below 
$T_\lambda$ \cite{Hodges03}, we infer that the sharp increase of $\lambda_Z$ occuring right 
below $T_\lambda$ is the signature of the building up of magnetic pair-correlations.
Down to the lowest temperature, the $\mu$SR spectra
have been recorded in the motional narrowing limit, since we compute
$\gamma_\mu\Delta_{\rm ZF} \, \tau_c \simeq 0.02 \ll 1$ at low temperature, 
taking as an estimate 
for $\tau_c$ the hyperfine field correlation time from M\"ossbauer 
($\sim 0.3$ ns) and computing $\Delta_{\rm ZF}$ $\simeq $ 0.08~T 
from the relation $\lambda_Z = 2 \gamma_\mu^2\Delta^2_{\rm ZF} \tau_c$. 
Therefore, the exponential form of the relaxation is fully justified.

An experimental estimate for $\tau_c$ can in principle be obtained
from the analysis of the field dependence of $\lambda_Z$ assuming that the 
properties of the system are not modified by an external field. In the case of
Yb$_3$Ga$_5$O$_{12}$ we have found that a field 
as low as 0.3 T has a strong influence, increasing rather than quenching 
$\lambda_Z$. We conclude that the field influences the system: this is not 
completely astonishing recalling that a field of 
0.7 T induces a phase transition for the isomorphous compound 
Gd$_3$Ga$_5$O$_{12}$ \cite{Schiffer94}.  

Yb$_3$Ga$_5$O$_{12}$ is therefore a second system with frustrated magnetic
interactions, after Yb$_2$Ti$_2$O$_7$, where no long-range order is found 
below $T_\lambda$. In the pyrochlore a sharp first-order transition appears in 
the fluctuation rate of the magnetic moments and below $T_\lambda$ they 
continue to fluctuate slowly (in the megahertz range) at a temperature 
independent frequency. In the garnet we have a new scenario: 
the fluctuations are still rapid below $T_\lambda$ and no abrupt change 
in their frequency is observed. Only dynamical short-range correlations 
build up. The continuous rise of $\lambda_Z$ is magnetic in origin, and 
therefore excludes a dimer phase to be formed below $T_\lambda$.

Among the different frustrated pyrochlores and garnets with a $\lambda$ peak
in SH that we have considered at the beginning of this letter, i.e. 
Gd$_2$Ti$_2$O$_7$, Gd$_2$Sn$_2$O$_7$, Er$_2$Ti$_2$O$_7$,
Yb$_2$Ti$_2$O$_7$, Dy$_3$Ga$_5$O$_{12}$ 
and Yb$_3$Ga$_5$O$_{12}$, some of them order and others do not. 
One can tentatively find a condition for a magnetic order to be present.
For all these compounds the variation of magnetic entropy at low temperature,
say below $\sim$ 10 K, is close to the expected value deduced from the 
number of electronic degrees of freedom. This is $R\ln 8$ for the Gd based 
compounds and $R\ln 2$ for the others. Inspecting Refs. 
\cite{Ramirez02,Bonville03,Blote69,Hodges02,Filippi80a,Filippi80b,Filippi81},
the fraction of the total entropy frozen at $T_\lambda$ is roughly
60~\% for Er$_2$Ti$_2$O$_7$ and Dy$_3$Ga$_5$O$_{12}$, 40 \%
for Gd$_2$Sn$_2$O$_7$ and 35 \% for Gd$_2$Ti$_2$O$_7$ whereas
it is only $\simeq$ 20 \% for Yb$_2$Ti$_2$O$_7$ and Yb$_3$Ga$_5$O$_{12}$. 
Therefore we estimate
that a long-range magnetic order is only present when at least $1/4-1/3$ 
of the magnetic entropy is released at $T_\lambda$.

While the entropy change at the $\lambda$-anomaly temperature is about the 
same for both Yb$_3$Ga$_5$O$_{12}$ and Yb$_2$Ti$_2$O$_7$, 
the Kramers ground-state doublet is approximately isotropic for the
former and strongly anisotropic for the latter. This suggests that a high 
magnetic anisotropy impedes the system from exploring different 
configurations, leading to an abrupt slowing down of the correlations.
$\Delta_{\rm ZF}$ is reduced by a factor $\simeq$ 14 \cite{Hodges02,Yaouanc03} 
when crossing $T_\lambda$
from above for Yb$_2$Ti$_2$O$_7$ whereas this reduction is much less 
($\alt$ 2) for Yb$_3$Ga$_5$O$_{12}$. This is again
consistent with the anisotropy difference between both systems: 
we expect $\Delta_{\rm ZF}$
to be small if the magnetic moments are confined to specific orientations.  

In conclusion, according to wisdom, a $\lambda$-type anomaly is indicative of 
a second order phase transition at $T_\lambda$ for a three-dimensional system. 
However, due to the presence of frustrated interactions, only dynamical 
short-range correlations build up below $T_\lambda$ in Yb$_3$Ga$_5$O$_{12}$. 
By comparison with results from other three-dimensional frustrated systems, 
we infer that the long-range 
order of the order parameter does not occur if a too large fraction of the 
entropy is released at a broad bump located above $T_\lambda$. The anisotropy 
of the system favors a freezing of the magnetic correlations.

We are grateful to J.A. Hodges for initiating this project and for his 
constant interest. We thank A. Forget for preparing the sample,
P. Bonville and J.A. Hodges for communication of their M\"ossbauer 
spectroscopy results prior to publication and useful discussions and
M.E. Zhitomirsky for a careful reading of the manuscript.
The authors from the 
Netherlands are grateful to the Dutch Scientific Organisation (NWO) for its 
financial support for the use of ISIS. Support from the European Community 
through its Access to Research Infrastructure action of the Improving Human 
Potential Program is acknowledged.

\bibliography{Yb3Ga5O12_paper.bib}

\begin{thebibliography}{33}
\bibitem[{Orb()}]{Orbachnote}
\bibinfo{note}{A Raman process can be theoretically effective but it does not
  fit the data}.
\end{thebibliography}

\end{document}